%% file: main.tex
\documentclass[aps, prl, twocolumn, nofootinbib, superscriptaddress]{revtex4-2}
\usepackage{amsmath}
\usepackage{amssymb}
\usepackage{graphicx}
\usepackage{blindtext}
\usepackage{mathrsfs}
\usepackage{color}
\usepackage{mwe}

\usepackage[colorlinks=true,
            linkcolor=blue,
            urlcolor=blue,
            citecolor=blue]{hyperref}

\newcommand{\GeV}{\mathrm{GeV}}

\begin{document}

\title{A New Gravitational Wave Power Law from Dissipative Sound Waves in Cosmological Phase Transitions}

\affiliation{School of Physics, Henan Normal University, Xinxiang 453007, P. R. China}

\author{Huai-Ke Guo}
\affiliation{International Centre for Theoretical Physics Asia-Pacific, University of Chinese Academy of Sciences, Beijing 100190, P. R. China}

\author{Sihao Xiao}
\email[]{Corresponding author, xiaosihao@itp.ac.cn}
\affiliation{Institute of Theoretical Physics, Chinese Academy of Sciences, Beijing 100190, P. R. China}
\affiliation{School of Physical Sciences, University of Chinese Academy of Sciences, Beijing 100049, P. R. China}

\author{Yang Xiao}
\email[]{Corresponding author, xiaoyangphy@gmail.com}
\affiliation{School of Physics, Henan Normal University, Xinxiang 453007, P. R. China}

\author{Jin Min Yang}
\affiliation{School of Physics, Henan Normal University, Xinxiang 453007, P. R. China}
\affiliation{Institute of Theoretical Physics, Chinese Academy of Sciences, Beijing 100190, P. R. China}

\author{Yang Zhang}
\affiliation{School of Physics, Henan Normal University, Xinxiang 453007, P. R. China}

\begin{abstract}

Power-law behavior in cosmological gravitational wave (GW) spectra reveals the physical nature of their sources. For sound waves generated by cosmological first-order phase transitions, we identify a new high-frequency power law arising from the dissipative evolution of the acoustic source. In a dissipative plasma, the GW spectrum acquires an additional $q^{-2}$ suppression relative to the perfect-fluid result in the late-time, strong-damping limit. The exponent is largely insensitive to the source profile, while the turnover and suppression strength probe plasma transport and the source lifetime. In particular, the turnover frequency in weakly coupled hidden sector plasmas is sensitive to the underlying gauge coupling and can lie in the millihertz band targeted by space-based interferometers, opening a way to probe microscopic plasma dynamics through the detailed GW spectrum.
\end{abstract}

\maketitle

\textbf{\textit{Introduction}}--Power-law behavior is one of the most informative features of stochastic gravitational wave (GW) spectra~\cite{Caprini:2009fx,Caprini:2018mtu,Kuroyanagi:2018csn,Cai:2019cdl}, as it can connect the observed spectral shape to the dynamics of the underlying sources. Various processes in the early Universe can produce characteristic power-law spectra. For instance, cosmological first-order phase transitions generate GWs through bubble collisions and the resulting acoustic motions~\cite{Hindmarsh:2013xza,Cutting:2018tjt}. Similar sound shells can also arise during primordial black hole formation and contribute to GW production~\cite{Ning:2025ogq}. Cosmic string networks generate broad GW spectra that carry information about string dynamics and the expansion history of the early Universe~\cite{Blanco-Pillado:2017oxo,Cui:2017ufi}, while the evolution and annihilation of domain wall networks can produce distinct power-law regimes~\cite{Hiramatsu:2013qaa,Notari:2025kqq}. These different sources motivate the use of spectral slopes and break frequencies to characterize their GW signals. In searches for stochastic GW backgrounds by the LIGO-Virgo-KAGRA collaborations~\cite{LIGOScientific:2026mjf,Virgo:2025aai} and pulsar timing array experiments~\cite{NANOGrav:2023gor,Xu:2023wog,Goncharov:2021oub,EPTA:2023fyk}, a small number of such parameters can provide compact templates for distinguishing different source scenarios and constraining their properties~\cite{Caprini:2019pxz,Gowling:2022pzb,Caprini:2024hue}. Beyond their phenomenological applications, power-law regimes are of particular theoretical interest because they can follow from general properties of the source, largely independent of its microscopic details~\cite{Caprini:2009fx,Cai:2019cdl}.

The acoustic GW spectrum from cosmological first-order phase transitions provides precisely such an example. Its characteristic spectral features are widely recognized to be primarily determined by the coarse-grained spatial scales of the source. Causality governs the infrared behavior, while the mean bubble separation and the sound-shell thickness set the peak and the characteristic breaks at higher frequencies~\cite{Hindmarsh:2016lnk,Hindmarsh:2017gnf,Hindmarsh:2019phv,Caprini:2009yp,RoperPol:2023dzg}. However, GW production is governed by unequal-time correlations of the anisotropic stress, indicating that the dynamical history of the source should also leave an imprint on the final spectrum~\cite{Caprini:2018mtu,Maggiore:2018sht}. Although previous studies have shown that the source history can modify the GW spectrum~\cite{Caprini:2009yp,Niksa:2018MHD,Auclair:2022Turbulence,Dahl:2024eup,Caprini:2024gyk}, isolating a distinct spectral scaling generated by this temporal evolution remains challenging. This raises a broader question: how is the dynamical history of the source encoded in the spectral structure of the resulting GWs?

Dissipative transport in a thermal plasma is a natural platform to explore this question. Such dissipation is an intrinsic feature of realistic plasmas, arising from the finite viscosity and mean free path of the plasma~\cite{Weinberg:1971mx,Kovtun:2012rj}. Related effects have been studied in scalar-induced GWs, where damping of scalar perturbations suppresses and reshapes the GW spectrum, leaving characteristic features associated with particle decoupling~\cite{Yu:2024xmz,Yu:2025cqu,Ekstedt:2025awx}. Dissipation can similarly modify the acoustic GW source generated by a first-order phase transition. These effects accumulate over the lifetime of the acoustic source, suppressing short-wavelength modes and reshaping the high-frequency GW spectrum~\cite{Guo:2023koq,Ellis:2020awk}. More importantly, dissipation may erase model-dependent details of the source and give rise to more robust features. The unavoidable presence of dissipation in realistic plasmas therefore calls for a systematic analytic treatment of its spectral corrections.

In this Letter, we identify for the first time an additional power law in the GW spectrum that originates from the temporal evolution of a dissipative acoustic source. In the late-time and strong-damping regime, this modification takes the form of a \(q^{-2}\) suppression relative to the perfect-fluid result. The scaling exponent is largely insensitive to the detailed acoustic source profile, while the suppression strength and turnover scale retain information about plasma transport. We further show that weakly coupled hidden-sector plasmas can generate sufficiently strong damping to make this power law modification potentially accessible to future space-based detectors.

\textbf{\textit{Evolution of a dissipative acoustic source}}--A power law generated by the temporal evolution of the source would provide a direct probe of the underlying dynamics. To isolate this contribution, we first consider the standard nondissipative case, where the GW spectrum of a stationary acoustic source can be written in the factorized form ~\cite{Xiao:2024rsj,Guo:2020grp}
\begin{equation}
	\mathcal{P}_{\text{GW}}^0(y,qL_f)=\tilde P_{\text{GW}}(qL_f)\,\Upsilon(y;w) y^{3w-1},
\end{equation}
where $qL_f$ is the dimensionless wave number and $y\equiv a/a_s$ tracks the evolution from the onset of the acoustic source. The factor $\widetilde P_{\rm GW}$ sets the source-dependent spectral shape, while $\Upsilon(y;w)$ encodes its finite-time growth in a background with equation-of-state parameter $w$. Since $\Upsilon$ is independent of $q$, the temporal evolution controls the overall growth but does not alter the spectral shape.

Dissipation breaks this simple factorization by introducing a momentum-dependent damping time, and the amplitude of each fluid velocity Fourier mode decays exponentially as
\small
\begin{equation}
    \tilde v^i_{\mathbf q}(\eta) \propto \exp\left[- \int_{\eta_*}^{\eta}\Gamma_q(\bar{\eta}) d\bar{\eta} \right] = \exp\left[- q^2d_{D}^2(\eta_{*},\eta) \right],
\end{equation} 
\normalsize
where $\eta_*$ is the reference conformal time from which dissipative damping begins to accumulate, and $\Gamma_q(\eta) \propto q^2$ is the damping rate, whose detailed form has been derived in Ref.~\cite{Weinberg:1971mx}. With this modification, the corresponding unequal-time correlator (UETC) of the velocity becomes 
\begin{equation}
\begin{aligned}
\label{eq:dD2stardefinition}
\left\langle
\tilde v_{\mathbf p}(\eta_1)
\tilde v_{\mathbf p}^{*}(\eta_2)
\right\rangle
&\sim
\mathcal P_v^{0}(pL_f)
\;\longrightarrow\;
\mathcal P_v(p;\eta_1,\eta_2),
\\
\mathcal P_v (p; \eta_1, \eta_2) &= \exp \left[ -p^2 d_D^2 (\eta_*, \eta_1, \eta_2) \right] \mathcal P_v^{(0)} (p)
\end{aligned}
\end{equation}
with $ d_D^2 (\eta_*, \eta_1, \eta_2) \equiv d_D^2 (\eta_*, \eta_1) + d_D^2 (\eta_*, \eta_2)$. Inserting Eq.~\eqref{eq:dD2stardefinition} into the UETC of the spatial energy-momentum tensor gives dissipatively corrected acoustic source spectrum
\small
\begin{equation}
	\label{eq:piUETCdiss}
    \begin{aligned}
        \Pi_{\text{diss}}(qL_f,q\eta_1,q\eta_2)
        &\propto
        \int d^3\tilde p\,
        \mathcal{K}(\eta_1-\eta_2, \tilde{p})
        \mathcal P_v^{(0)}(\tilde p)\mathcal P_v^{(0)}(\tilde{\bar p})
        \\
        &\quad \times
        \exp[-(\tilde{p}^2+\tilde{\bar p}^2)d_D^2(\eta_*,\eta_1,\eta_2)/L_f^2],
    \end{aligned}
\end{equation} 
\normalsize
where $\tilde p=pL_f$ and $\tilde{\bar p}=\bar pL_f$ are the dimensionless momenta, with $\bar p=|\mathbf p-\mathbf q|$. The kernel $\mathcal K$ contains the angular and oscillatory structure of the nondissipative UETC.
Equation~\eqref{eq:piUETCdiss} shows that dissipation couples the temporal evolution to the internal momenta through the exponential damping factor. As a result, different momentum modes are reweighted during the source evolution, leading to a modification of the GW spectral shape. To obtain a simple analytic description of this behavior, we introduce two approximations:

(i) \textbf{Frozen-damping approximation}. The damping rate $\Gamma_q(q,\eta)=q^2D_a(\eta)$ is generally time dependent, mainly through the transport coefficient $D_a(\eta)$. For a nearly conformal plasma in radiation domination, $D_a\propto(\eta_{\rm sh}/s)/(aT)$, where $\eta_{\rm sh}$ and $s$ denote the shear viscosity and entropy density of the plasma, respectively~\cite{Weinberg:1971mx,Kovtun:2012rj}. Since $aT$ is approximately constant and $\eta_{\rm sh}/s$ varies only slowly with the thermal background, $D_a$ changes mildly during the relevant acoustic stage. We therefore approximate it by its value at a characteristic time, $D_a(\eta)\simeq D_{a,*}$. With this approximation, the damping exponent in Eq.~\eqref{eq:piUETCdiss} can be characterized by the dimensionless parameter $\hat{\delta}_D\equiv 2D_{a,*}/(a_sH_sL_f^2)$, which measures the accumulated diffusion over one Hubble time in units of $L_f^2$.

(ii) \textbf{Characteristic-momentum approximation.} 
For a fixed GW wave vector $\mathbf q$, momentum conservation links the two sound modes with momenta $\mathbf p$ and $\mathbf p-\mathbf q$. Since the velocity spectrum has a single maximum, the product $P_v^{(0)}(p)P_v^{(0)}(|\mathbf p-\mathbf q|)$ restricts the dominant contribution to a limited momentum range. Dissipation sharpens this selection because $p^2+|\mathbf p-\mathbf q|^2= \frac{q^2}{2} + 2\left|\mathbf p-\frac{\mathbf q}{2}\right|^2$,
which favors nearly symmetric momentum sharing. The kernel $\mathcal K$, however, vanishes at the exactly symmetric point and shifts the dominant contribution to a nearby interior region. These effects therefore define a characteristic momentum pair, and we make the replacement $\tilde p^2+\tilde{\bar p}^{\,2} \to S_*(qL_f) \equiv \tilde p_*^2(qL_f)+\tilde{\bar p}_*^2(qL_f)$. This approximation isolates the dissipative time dependence from the momentum convolution. Its accuracy is tested against the full result in the Supplemental Material.


With the above two approximations, the conformal time dependence of the dissipative UETC can be pulled out of the momentum integral, allowing the dissipative GW spectrum to be factorized as
\begin{equation}
    \mathcal P_{\rm GW,diss}(y,qL_f)
    \simeq
    \widetilde P_{\rm GW}(qL_f)\,
    \Upsilon_{\rm diss}(y,q;w) y^{3w-1},
\end{equation}
where
\small
\begin{equation}
		\Upsilon_{\rm diss}(y,q;w)
		= \mathcal A_w(q)\,
		e^{\delta_D(q)y_{+,*}}
		\left[
		\Gamma\!\left(s_w,x_q\right)
		-\Gamma\!\left(s_w,x_q y^{\alpha_w}\right)
		\right],
	\label{eq:Upsilondissipation}
\end{equation}
\normalsize
where $\delta_D(q)\equiv\hat\delta_D S_*(qL_f)$,  while the remaining quantities are defined in the Supplemental Material.

\begin{figure}[htb]
    \centering
    \includegraphics[width=0.49\textwidth]{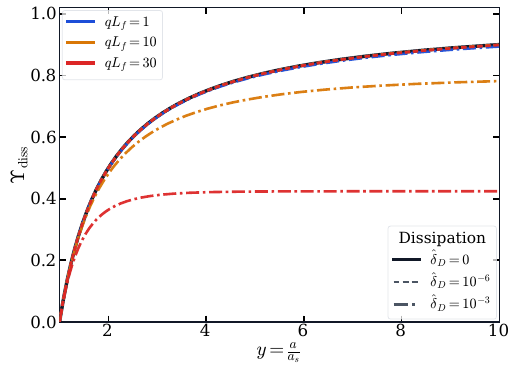}
    \vspace*{-.7cm}
    \caption{ Dissipative growth factor $\Upsilon_{\rm diss}$ versus $y=a/a_s$ for $qL_f=\{1,10,30\}$ and $\hat{\delta}_D=\{0,10^{-6},10^{-3}\}$, assuming a radiation-dominated background with $w=1/3$. Colors label $qL_f$ and line styles label $\hat{\delta}_D$, where $\hat{\delta}_D=0$ gives the nondissipative limit.}
    \label{fig:UpsilonDiss}
\end{figure}

The behavior of $\Upsilon_{\rm diss}$ for a radiation-dominated background is shown in Fig.~\ref{fig:UpsilonDiss}. Unlike the nondissipative $\Upsilon$ factor, $\Upsilon_{\rm diss}$ depends explicitly on the wave number through $\delta_D(q)$. For small $\hat{\delta}_D$, the result is almost indistinguishable from the nondissipative case, while visible deviations appear for larger dissipation. 

\textbf{\textit{Dissipation-induced power law}}--The explicit wave-number dependence of $\Upsilon_{\rm diss}$ implies that dissipation modifies not only the accumulated GW amplitude but also the spectral shape. Within this factorized description, the dissipative correction is entirely encoded in $\Upsilon_{\rm diss}$, while the source-dependent factor $\widetilde P_{\rm GW}$ remains unchanged. The relative modification of the GW spectrum is therefore given by
\begin{equation}
\label{eq:upsilon_over_upsilon}
\mathcal{R}_D = \frac{\mathcal P_{\rm GW,diss}}{\mathcal P_{\rm GW}^{0}}=\frac{\Upsilon_{\rm diss}}{\Upsilon}.
\end{equation}

For $w<1$, in the late-time and strong-damping limit, the dissipative factor asymptotes to $\Upsilon_{\rm diss}^{\infty}\simeq1/[\hat{\delta}_D S_*(qL_f)]$ and the nondissipative factor behaves as $\Upsilon_\infty(w)=2 / \left[3(1-w)\right]$. One therefore obtains
\begin{align}
    \mathcal{R}_D^{\infty}(q;w)
    &\sim
    \frac{3(1-w)}{2}
    \frac{1}{\hat{\delta}_D S_*(qL_f)} .
\end{align}

In the large $qL_f$ limit, the peak momentum satisfies $\tilde p_*\simeq qL_f$, so that $S_*(qL_f)\equiv \tilde p_*^{\,2}+\tilde{\bar p}_*^{\,2}\simeq(qL_f)^2+\tilde{\bar p}_*^{\,2}\sim(qL_f)^2$. The acoustic GW spectrum therefore receives the asymptotic correction
\begin{equation}
    \frac{\mathcal{P}_{\text{GW,diss}}(y,qL_f)}
         {\mathcal{P}^0_{\text{GW}}(y,qL_f)}
    \sim
    \frac{3(1-w)}{2\hat{\delta}_D(qL_f)^2} .
\end{equation}

Although the preceding result was obtained using the characteristic-momentum approximation, the additional $q^{-2}$ suppression law does not in fact rely on this approximation. In the late-time strong-damping regime, the full momentum integral is controlled by resonant sound-mode pairs satisfying $qL_f=c_s(\tilde p+\tilde{\bar p})$, as shown in the Supplemental Material. On this resonant surface, $S_*\equiv\tilde p^{\,2}+\tilde{\bar p}^{\,2}$ is restricted to $(qL_f)^2/(2c_s^2)\leq S_*\leq(qL_f)^2/c_s^2$. Different source models only reweight these pairs, changing the bounded prefactor but not the $q^2$ scaling. As dissipation accumulates over the source lifetime, it limits the effective duration of GW production to the inverse damping rate, yielding the additional $q^{-2}$ suppression after the full momentum integration.

The emergence of this regime requires sufficiently strong damping over the acoustic lifetime. It is therefore important to determine whether the required dissipation strength can be realized in a thermal plasma. For a nearly conformal relativistic plasma in a radiation-dominated universe, $\hat{\delta}_D$ can be estimated in terms of the shear viscosity $\eta_{\rm sh}$ and bulk viscosity $\zeta$.  Using
$R_*\simeq(8\pi)^{1/3}v_w/\beta$ and the definition of $\hat\delta_D$, we obtain
\small
\begin{align}
	\hat{\delta}_D&\simeq 4.8\times10^{-14}
	\left(\frac{\eta_{\rm sh}/s}{1/(4\pi)}\right)
	\left(1+\frac{3\zeta}{4\eta_{\rm sh}}\right)
	\left(\frac{\beta/H_*}{100}\right)^2
	\notag\\
	&\quad\times
	\left(\frac{0.6}{v_w}\right)^2
	\left(\frac{T_*}{100\,\GeV}\right)
	\left(\frac{g_*}{106.75}\right)^{1/2}
	\left(\frac{R_*}{L_f}\right)^2,
	\label{eq:hatdelta_estimate_main}
\end{align}
\normalsize
where $\beta$ is the inverse duration of the transition, $H_*$ is the Hubble rate at $T_*$, $v_w$ is the wall velocity, and $R_*$ is the mean bubble separation. With these reference values, an electroweak-scale transition in an ordinary plasma is typically only weakly damped. A larger $\hat{\delta}_D$ can arise  when the acoustic source is dominated by a deconfined, weakly coupled hidden non-Abelian plasma. At leading logarithmic order, its
shear viscosity scales as
\begin{equation}
	\frac{\eta_X}{s_X}
	\simeq
	\frac{\mathcal C_\eta}
	{g_X^4\ln(1/g_X)},
\end{equation}
where $g_X\equiv g_X(\mu\sim T_X)$ is the running hidden gauge coupling
and $\mathcal C_\eta\simeq 3.9$ for a pure $SU(3)$ plasma~\cite{Arnold:2000dr,Arnold:2003zc}. Taking $T_X=T_*$ for illustration and neglecting the bulk viscosity,
Eq.~\eqref{eq:hatdelta_estimate_main} gives
\small
\begin{align}
	\hat{\delta}_D
	&\simeq
	7.1\times10^{-4}
	\left(\frac{\mathcal C_\eta}{3.9}\right)
	\left(\frac{0.005}{g_X}\right)^4
	\frac{\ln 200}{\ln(1/g_X)}
	\left(\frac{\beta/H_*}{100}\right)^2
	\notag\\
	&\quad\times
	\left(\frac{0.6}{v_w}\right)^2
	\left(\frac{T_*}{100\,\GeV}\right)
	\left(\frac{g_*}{106.75}\right)^{1/2}
	\left(\frac{R_*}{L_f}\right)^2 .
	\label{eq:hatdelta_hidden_plasma}
\end{align}
\normalsize
For the reference transition parameters and $g_X\simeq5\times10^{-3}$,
 $\hat{\delta}_D$ can reach $10^{-3}$. The inclusion of light fermionic ($m_f\ll T_X$) degrees of freedom can further increase $\mathcal C_\eta$, thereby allowing $\hat{\delta}_D\sim10^{-3}$ to be reached with less restrictive parameter choices~\cite{Arnold:2000dr}. We therefore use \(\hat\delta_D=10^{-3}\) as the benchmark in the numerical spectrum below.

\begin{figure*}[t]
    \centering
    \includegraphics[width=\textwidth]{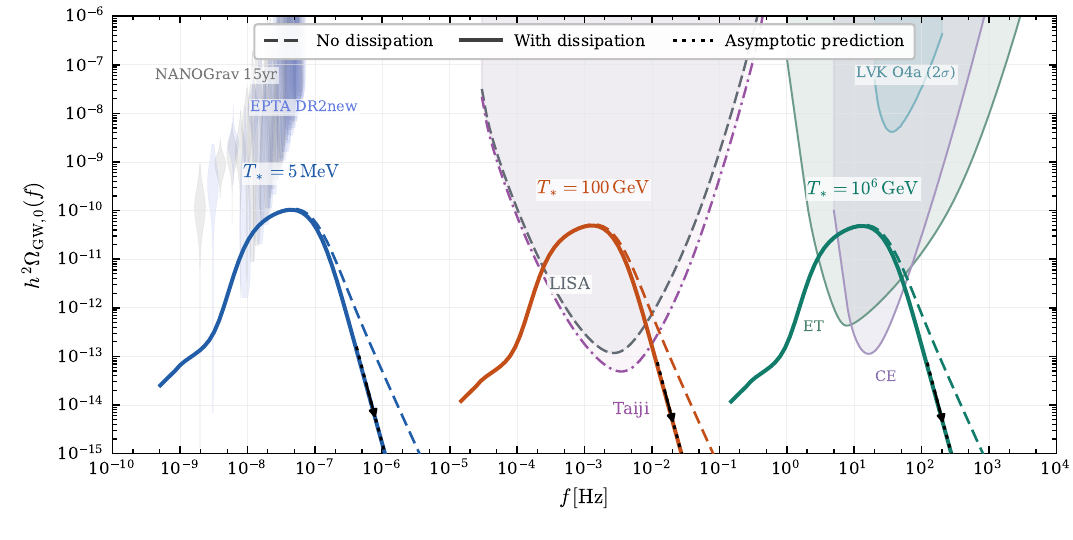}
    \vspace*{-.9cm}
    \caption{Present-day acoustic GW spectra for \(T_*=\{5\,\mathrm{MeV},100\,\mathrm{GeV},10^6\,\mathrm{GeV}\}\) (blue, orange, and green). Colored dashed and solid curves show nondissipative and dissipative (\(\hat\delta_D=10^{-3}\)) results; black dotted segments with arrows show the asymptotic additional \(q^{-2}\) suppression. We fix \(\alpha_N=0.3\), \(v_w=0.6\), \(\beta/H_*=100\), \(L_f=R_*\), \(y_{\rm final}/y_s=2\), and \(g_*=\{10.75,100,106.75\}\), adjusting transport coefficients to maintain \(\hat\delta_D\). Thin curves show projected LISA~\cite{amaro2017laser}, Taiji~\cite{10.1093/ptep/ptaa083}, ET~\cite{ET:2025xjr}, and CE~\cite{Reitze:2019iox} power-law-integrated sensitivities (4 effective years, SNR\(=10\); \(A+E\) channels for LISA and Taiji). PTA regions show NANOGrav 15-year~\cite{NANOGrav:2023gor} and EPTA DR2new~\cite{EPTA:2023fyk} free-spectrum posteriors, while LVK O4a denotes the current \(2\sigma\) power-law sensitivity~\cite{Virgo:2025aai}.}
    \label{fig:PGWdissTail}
\end{figure*}


To test this prediction, we compute the full acoustic GW spectrum within the improved sound shell model~\cite{RoperPol:2023dzg}. As shown in Fig.~\ref{fig:PGWdissTail}, dissipation leaves the spectrum nearly unchanged around the peak but increasingly suppresses the high-frequency tail. For the electroweak-scale benchmark considered here (orange line), a fit to the suppression ratio $\mathcal R_D\equiv\mathcal P_{\rm GW,diss}/\mathcal P_{\rm GW}^{0}$ gives $\mathcal R_D\propto f^{-1.965}$ for $f\gtrsim7.5\times10^{-2}~\mathrm{Hz}$, differing from the predicted additional $f^{-2}$ scaling by less than $2\%$. The black dotted curve, obtained from the asymptotic prediction, closely tracks the full dissipative spectrum in the ultraviolet. The overall coefficient $\hat{\delta}_D(qL_f)^2\mathcal{R}_D$ remains source dependent but is bounded: for the radiation-dominated benchmark and source duration considered here, the Supplemental Material gives a range from $2/3$ to $4/3$. The numerical value, about $1.2$--$1.3$, lies within this interval. Thus the source profile changes the strength of the suppression only within a restricted range, while the additional $f^{-2}$ scaling remains robust.

\begin{figure}[t]
    \centering
    \includegraphics[width=\linewidth]{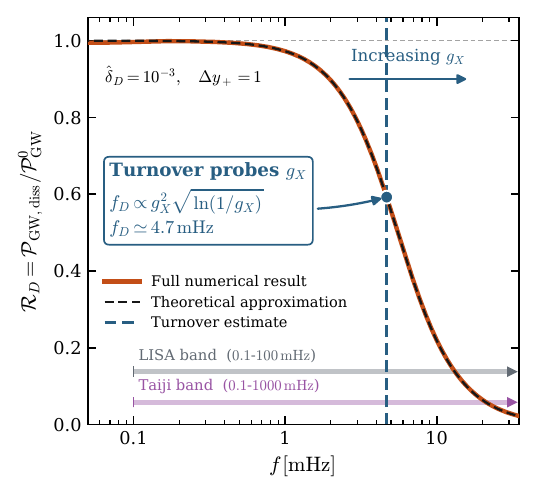}
    \caption{Dissipative turnover for the electroweak-scale source of
    Fig.~\ref{fig:PGWdissTail}. The orange solid and black dashed curves
    show the full numerical result and theoretical approximation.
    The blue marker denotes $f_D\simeq4.7\,\mathrm{mHz}$, which increases
    with the hidden gauge coupling $g_X$ at fixed source parameters.
    Gray and purple strips indicate the nominal
    LISA~\cite{amaro2017laser} and
    Taiji~\cite{10.1093/ptep/ptaa083} frequency bands, respectively.}
    \label{fig:dissipativeTurnover}
\end{figure}

The asymptotic power-law regime develops only after damping has become efficient, where the GW signal may already be strongly suppressed. A more accessible signature is therefore the turnover at which the dissipative spectrum first departs from the perfect-fluid prediction. This scale is determined by $\hat{\delta}_D S_*[(qL_f)_D]\Delta y_+ \sim 1$, where $\Delta y_+ \equiv a_s H_s(\eta_f-\eta_s)$ denotes the dimensionless duration of the acoustic source. Physically, this condition corresponds to the damping time of the relevant sound modes becoming comparable to the acoustic source lifetime. Unlike the conventional acoustic breaks set by the mean bubble separation and the sound-shell thickness, the dissipative turnover is controlled by the source lifetime and the transport properties of the plasma. For the electroweak-scale benchmark point shown in Fig.~\ref{fig:dissipativeTurnover}, $\Delta y_+=1$ and we find $(qL_f)_D\simeq32$, corresponding to $f_D\simeq4.7\,{\rm mHz}$. The onset of the dissipative modification therefore lies in the frequency range targeted by space-based interferometers. The high-frequency slope and the turnover carry complementary information: the former reflects the momentum dependence of the damping rate, while the latter determines the scale at which the dissipative dynamics becomes important.

More broadly, the dissipative turnover opens a route to extracting microscopic information from a phase-transition GW spectrum, rather than using the signal mainly as a test of detectability. Its position depends on $\hat{\delta}_D$ and therefore retains information about the transport properties of the primordial plasma. In the weakly coupled hidden-sector example considered above, $\hat{\delta}_D\propto[g_X^4\ln(1/g_X)]^{-1}$, so the turnover is also sensitive to the underlying gauge interaction. The detailed GW spectrum can therefore carry information about plasma dynamics beyond the Standard Model that is not contained in the overall amplitude or peak frequency alone. This extends the role of phase-transition GWs from identifying a possible early-Universe transition toward probing the microscopic properties of the plasma that generated the signal. A quantitative extraction of the microscopic parameters from the GW spectrum is left for future work.


\textbf{\textit{Conclusion}}--
In this work, we showed that dissipation can leave a distinct imprint on the acoustic GW spectrum from a cosmological first-order phase transition. Although the sound waves themselves are exponentially damped, the GW spectrum develops an additional power-law suppression rather than an exponential cutoff. For diffusive damping, this gives an additional $q^{-2}$ scaling at high frequencies. The result is largely insensitive to the detailed acoustic source profile and is confirmed by the numerical results.

The main implication is that dissipation does more than reduce the GW amplitude. It leaves a distinct structure in the spectrum that carries information about the time evolution of the source. The high-frequency slope reflects the momentum dependence of the damping rate, while the turnover marks the scale at which dissipation becomes important. These features therefore provide information that complements the usual peak and acoustic break scales. This opens the possibility of using the detailed GW spectrum to probe transport properties and microscopic interactions in the primordial plasma. More generally, a damping law $\Gamma_q\propto q^n$ leads to an additional $q^{-n}$ suppression relative to the perfect-fluid GW spectrum, suggesting a broader connection between the spectral structure of GWs and the relaxation dynamics of their sources. 

Experimentally, these new spectra represent potential signals hiding in the current observational data of the LIGO-Virgo-KAGRA collaborations, pulsar timing array experiments, and our results call for extending the currently used templates to include the dissipative turnover and the modified high-frequency power law.

\textbf{\textit{Acknowledgments}}--
This work was supported by the National Natural Science Foundation of China (Grant Nos. 12335005, 12475109, 12547104, and 12605177), the startup fund provided by the University of Chinese Academy of Sciences, the PI Research Fund of Henan Normal University (Grant No. 5101029470335), the Natural Science Foundation of Henan Province (Grant No. 262300421233), and the Youth Project of the Key Scientific Research Project of Colleges and Universities in Henan Province (Grant No. 27AQ140008).
\bibliographystyle{apsrev4-2}
\bibliography{reference}

\clearpage
\onecolumngrid

\setcounter{equation}{0}
\setcounter{figure}{0}
\setcounter{table}{0}
\renewcommand{\theequation}{S\arabic{equation}}
\renewcommand{\thefigure}{S\arabic{figure}}
\renewcommand{\thetable}{S\arabic{table}}

\renewcommand{\theHequation}{S\arabic{equation}}
\renewcommand{\theHfigure}{S\arabic{figure}}
\renewcommand{\theHtable}{S\arabic{table}}

\begin{center}
    \textbf{Supplemental Material}
\end{center}

\input{supplemental_material}

\end{document}

%% file: supplemental_material.tex
This Supplemental Material provides the derivations supporting the main text.
We first obtain the dissipative growth factor and its limiting behavior, then
examine the characteristic-momentum approximation and derive the high-frequency
$q^{-2}$ scaling without relying on it. Unless stated otherwise, the notation
and benchmark choices follow those of the main text.

\section{Derivation of the dissipative growth factor}
\label{appendix:Upsilon}

We derive the dissipative growth factor in
Eq.~(6) of the main text, keeping the source normalization
explicit throughout. We consider tensor perturbations around a spatially flat FLRW metric,
\begin{equation}
    ds^2=a^2(\eta)
    \left[
    -d\eta^2+
    \left(\delta_{ij}+h_{ij}\right)dx^i dx^j
    \right],
    \label{eq:app_metric}
\end{equation}
where $h_{ij}$ is transverse and traceless. In Fourier space, each tensor mode
obeys
\begin{equation}
    h_{ij}''+2\mathcal H h_{ij}'+q^2h_{ij}
    =
    16\pi G a^2\pi_{ij}^{\rm TT},
    \label{eq:app_tensor_eq}
\end{equation}
where $q\equiv|\mathbf q|$ is the comoving wave number,
$\mathcal H=a'/a$, and primes denote conformal-time derivatives. The source
$\pi_{ij}^{\rm TT}$ follows from the energy-momentum tensor of the acoustic
fluid. We denote its homogeneous, spatially averaged energy density and
pressure by $\bar e_f(\eta)$ and $\bar p_f(\eta)$ and define the mean enthalpy
density as
\begin{equation}
    \bar w_f(\eta)\equiv\bar e_f(\eta)+\bar p_f(\eta).
    \label{eq:app_mean_enthalpy}
\end{equation}
The subscript $f$ distinguishes the fluid quantities from the background
equation-of-state parameter $w$. For an ideal fluid, the energy-momentum tensor
is
\begin{equation}
    T_{\mu\nu}
    =
    (e_f+p_f)u_\mu u_\nu+p_f g_{\mu\nu},
    \label{eq:app_fluid_emt}
\end{equation}
where $e_f$ and $p_f$ are the local thermodynamic fields and $u^\mu$ is the
fluid four-velocity. Writing $e_f=\bar e_f+\delta e_f$ and
$p_f=\bar p_f+\delta p_f$, the isotropic pressure term vanishes under the TT
projection, while enthalpy fluctuations multiplying $v_i v_j$ enter only at
higher order. The leading acoustic stress is therefore the term quadratic in
the fluid velocity,
\begin{equation}
    T_{ij}^{(v)}(\eta,\mathbf q)
    \simeq
    a^2(\eta)\bar w_f(\eta)
    \int\frac{d^3p}{(2\pi)^3}
    \tilde v_i(\eta,\mathbf p)
    \tilde v_j(\eta,\mathbf q-\mathbf p),
    \label{eq:app_stress_velocity}
\end{equation}
where $\tilde v_i(\eta,\mathbf p)$ is the spatial Fourier mode of the fluid
velocity. The transverse-traceless part of the spatial stress is
\begin{equation}
    \pi_{ij}^{\rm TT}(\eta,\mathbf q)
    =
    a^{-2}(\eta)\Lambda_{ij,lm}(\hat{\mathbf q})
    T_{lm}^{(v)}(\eta,\mathbf q),
    \qquad
    \Lambda_{ij,lm}
    =
    P_{il}P_{jm}-\frac12P_{ij}P_{lm},
    \label{eq:app_TT_projector}
\end{equation}
where $P_{ij}=\delta_{ij}-\hat q_i\hat q_j$. Using
Eq.~\eqref{eq:app_stress_velocity}, the TT source becomes
\begin{equation}
    \pi_{ij}^{\rm TT}(\eta,\mathbf q)
    \simeq
    \bar w_f(\eta)\Lambda_{ij,lm}(\hat{\mathbf q})
    \int\frac{d^3p}{(2\pi)^3}
    \tilde v_l(\eta,\mathbf p)
    \tilde v_m(\eta,\mathbf q-\mathbf p).
    \label{eq:app_acoustic_TT_source}
\end{equation}
We define its unequal-time correlator by
\begin{equation}
    \left\langle
    \pi_{ij}^{\rm TT}(\eta_1,\mathbf q_1)
    \pi_{ij}^{\rm TT}(\eta_2,\mathbf q_2)
    \right\rangle
    =
    (2\pi)^3\delta^3(\mathbf q_1+\mathbf q_2)
    \Pi(q_1,\eta_1,\eta_2).
    \label{eq:app_source_uetc}
\end{equation}
Because the source in Eq.~\eqref{eq:app_acoustic_TT_source} is quadratic in the
velocity, its UETC is determined by the velocity two-point function. For the
statistically isotropic, longitudinal flow used in the sound-shell
model, we define the nondissipative dimensionless
spectrum $\mathcal P_v^0$ by
\begin{equation}
\begin{aligned}
    \left\langle
    \tilde v^{i(0)}_{\mathbf p}(\eta_1)
    \tilde v^{j(0)*}_{\mathbf k}(\eta_2)
    \right\rangle
    &=(2\pi)^3\delta^3(\mathbf p-\mathbf k)
    \hat p^i\hat p^j
    \frac{2\pi^2}{p^3}
    \mathcal P_v^0(pL_f)
    \\
    &\quad\times
    \cos\left[c_sp(\eta_1-\eta_2)\right],
\end{aligned}
    \label{eq:app_velocity_correlator}
\end{equation}
where $p\equiv|\mathbf p|$, $L_f$ is the characteristic comoving acoustic
length, $c_s$ is the sound speed, and the superscript $(0)$ labels a
nondissipative mode. Using Eq.~\eqref{eq:app_velocity_correlator}, the source UETC can be
written as
\begin{equation}
\begin{aligned}
    \Pi_0(q,\eta_1,\eta_2)
    &=
    \frac{a_s^8L_f^3}{a^4(\eta_1)a^4(\eta_2)}
    \left[
    \bar w_{f,s}\bar U_f^2
    \right]^2
    \\
    &\quad\times
    \tilde\Pi_0(qL_f,q\eta_1,q\eta_2),
\end{aligned}
    \label{eq:app_dimensionless_uetc}
\end{equation}
Here $\bar U_f$ is the rms velocity of the nondissipative fluid, and $\eta_s$
denotes the effective onset of the acoustic source. We define
$a_s\equiv a(\eta_s)$ and $\bar w_{f,s}\equiv\bar w_f(\eta_s)$. In writing
Eq.~\eqref{eq:app_dimensionless_uetc}, we have used
$\bar w_f(\eta)=\bar w_{f,s}[a_s/a(\eta)]^4$, appropriate for a radiation-like
acoustic fluid. The subscript $0$ denotes the nondissipative source. With
$\mathbf Q\equiv L_f\mathbf q$, $Q\equiv|\mathbf Q|=qL_f$,
$\tilde{\mathbf p}\equiv L_f\mathbf p$,
$\tilde p\equiv|\tilde{\mathbf p}|$,
$\bar p\equiv|\mathbf p-\mathbf q|$, $\tilde{\bar p}\equiv\bar pL_f$, and
$\mu\equiv\hat{\mathbf q}\cdot\hat{\mathbf p}$, the dimensionless UETC is
\begin{equation}
\begin{aligned}
    \tilde\Pi_0(Q,q\eta_1,q\eta_2)
    &=
    \frac{\pi}{2\bar U_f^4}
    \int d^3\tilde p\,
    \frac{(1-\mu^2)^2}{\tilde p\,\tilde{\bar p}^{5}}
    \mathcal P_v^0(\tilde p)
    \mathcal P_v^0(\tilde{\bar p})
    \\
    &\quad\times
    \cos[c_sp(\eta_1-\eta_2)]
    \cos[c_s\bar p(\eta_1-\eta_2)].
\end{aligned}
    \label{eq:app_sound_shell_uetc}
\end{equation}

With the source normalization and momentum dependence separated, we describe
the finite source duration through the scale-factor ratio
\begin{equation}
    y(\eta)\equiv\frac{a(\eta)}{a_s},
    \qquad
    H_s\equiv H(\eta_s)=\frac{\mathcal H_s}{a_s}
    =\frac{a'(\eta_s)}{a_s^2},
    \label{eq:app_y_definition}
\end{equation}
where $H_s$ is the physical Hubble rate at $\eta_s$. It follows that
$y(\eta_s)=1$ and
$\bar w_f(\eta)=y^{-4}(\eta)\bar w_{f,s}$. This scaling concerns the acoustic
source and is distinct from the background equation of state that controls the
expansion. For a constant background relation $p_{\rm bg}=w\rho_{\rm bg}$ with
$w>-1/3$, the scale factor obeys
\begin{equation}
    y^{(1+3w)/2}
    =
    1+\frac{1+3w}{2}a_sH_s(\eta-\eta_s).
    \label{eq:app_y_eta_relation}
\end{equation}
To separate the time difference relevant to source stationarity from the mean
time governing the background evolution, we introduce $y_-$ and $y_+$. Writing
$y_i\equiv y(\eta_i)$ and $\eta_+\equiv(\eta_1+\eta_2)/2$, we define
\begin{equation}
\begin{aligned}
    y_-
    &\equiv
    \frac{2}{1+3w}
    \left(y_1^{(1+3w)/2}-y_2^{(1+3w)/2}\right),
    \\
    y_+
    &\equiv
    \frac{1}{1+3w}
    \left(y_1^{(1+3w)/2}+y_2^{(1+3w)/2}\right).
\end{aligned}
    \label{eq:app_yplus_yminus_definition}
\end{equation}
Equation~\eqref{eq:app_y_eta_relation} then gives
\begin{equation}
    y_-=a_sH_s(\eta_1-\eta_2),
    \qquad
    y_+=\frac{2}{1+3w}+a_sH_s(\eta_+-\eta_s).
    \label{eq:app_yplus_yminus_eta}
\end{equation}
These relations show that $y_-$ is proportional to the time difference
$\eta_1-\eta_2$, whereas $y_+$ tracks the mean time $\eta_+$.
For a stationary nondissipative source, the UETC therefore depends on $y_-$ but
not on $y_+$. With $\tilde q\equiv q/(a_sH_s)$, this gives
\begin{equation}
    \tilde\Pi_0(qL_f,q\eta_1,q\eta_2)
    =
    \tilde\Pi_0(Q,q(\eta_1-\eta_2))
    =
    \tilde\Pi_0(Q,\tilde q\,y_-).
    \label{eq:app_nondiss_uetc_stationary}
\end{equation}

Combining this stationary UETC with the retarded solution of
Eq.~\eqref{eq:app_tensor_eq} and the standard expression for the GW energy
density gives
\begin{equation}
\begin{aligned}
    \mathcal P_{\rm GW}^{0}(y,Q)
    &\simeq
    \frac{\left(16\pi G\bar w_{f,s}\bar U_f^2\right)^2
    Q^3}{24\pi^2H^2H_s^2y^4}
    \\
    &\quad\times
    \int_{\mathcal D}dy_+dy_-
    \frac{\mathcal G(\tilde y,\tilde y_1,\tilde y_2)}
    {y_1^{3(1+w)/2}y_2^{3(1+w)/2}}
    \frac{\tilde\Pi_0(Q,\tilde q\,y_-)}{\tilde q^2},
\end{aligned}
    \label{eq:app_gw_double_time}
\end{equation}
where $H=H(y)$, $\tilde y\equiv y\tilde q^{\,2/(1+3w)}$,
$\tilde y_i\equiv y_i\tilde q^{\,2/(1+3w)}$, and $\mathcal G$ is the
dimensionless Green-function combination. The domain $\mathcal D$ denotes the
physical double-time integration region.

The two time variables now play different roles: the full relative-time
kernel, combining GW propagation with the source UETC, oscillates rapidly with
$y_-$, while its envelope and the cosmological background vary slowly with
$y_+$.
For large $\tilde q$ and $y_+\gg|y_-|$, the $y_-$ integral defines the
source-dependent spectrum $\tilde P_{\rm GW}(Q)$. In the same limit, the
large-$\tilde q$ asymptotic form of the Green function, together with
$y_1\simeq y_2$, reduces the slowly varying kernel to
$c_w^{\gamma_w}y_+^{-\gamma_w}$. Since the double-time domain corresponds to
$c_w\leq y_+\leq c_wy^{\alpha_w}$, the spectrum factorizes as
\begin{equation}
    \mathcal P_{\rm GW}^{0}(y,Q)
    =
    \tilde P_{\rm GW}(Q)\Upsilon(y;w)y^{3w-1},
    \label{eq:app_nondiss_factorized}
\end{equation}
with
\begin{equation}
    \Upsilon(y;w)
    =
    c_w^{\gamma_w}
    \int_{c_w}^{c_wy^{\alpha_w}}
    dy_+\,y_+^{-\gamma_w},
    \label{eq:app_upsilon_nondiss_integral}
\end{equation}
where
\begin{equation}
    c_w\equiv \frac{2}{1+3w},
    \qquad
    \gamma_w\equiv \frac{4}{1+3w},
    \qquad
    \alpha_w\equiv \frac{1+3w}{2}.
    \label{eq:app_c_gamma_alpha}
\end{equation}
The remaining integral can be evaluated directly, giving
\begin{equation}
    \Upsilon(y;w)
    =
    \frac{2\left[y^{3(w-1)/2}-1\right]}
    {3(w-1)},
    \qquad w\neq1,
    \label{eq:app_upsilon_nondiss}
\end{equation}
in agreement with the main text.

With the nondissipative factorization established, we now include damping of
the acoustic velocity modes. In this case, the velocity mode 
evolves as
\begin{equation}
    \tilde v^i_{\mathbf p}(\eta)
    =
    \exp\left[-p^2d_D^2(\eta_*,\eta)\right]
    \tilde v^{i(0)}_{\mathbf p}(\eta),
    \label{eq:app_damped_velocity_mode}
\end{equation}
where $\eta_*$ is the time from which damping accumulates and
$\tilde v^{i(0)}_{\mathbf p}$ is the corresponding nondissipative mode. As in
the main text, we take $\eta_*=\eta_s$, so damping starts with the acoustic
source; we retain $\eta_*$ in the intermediate expressions only to display its
role. The accumulated diffusion length squared is
\begin{equation}
    d_D^2(\eta_a,\eta_b)
    \equiv
    \int_{\eta_a}^{\eta_b}D_a(\bar\eta)d\bar\eta.
    \label{eq:app_dD_definition}
\end{equation}
Here $D_a$ is the amplitude-damping coefficient in conformal time. Substituting Eq.~\eqref{eq:app_damped_velocity_mode} into
Eq.~\eqref{eq:app_velocity_correlator} gives the dissipative two-time
correlator
\begin{equation}
\begin{aligned}
    \left\langle
    \tilde v^i_{\mathbf p}(\eta_1)
    \tilde v^{j*}_{\mathbf k}(\eta_2)
    \right\rangle
    &=(2\pi)^3\delta^3(\mathbf p-\mathbf k)
    \hat p^i\hat p^j
    \frac{2\pi^2}{p^3}
    \mathcal P_v(p;\eta_1,\eta_2)
    \\
    &\quad\times
    \cos\left[c_sp(\eta_1-\eta_2)\right],
\end{aligned}
    \label{eq:app_diss_velocity_correlator}
\end{equation}
where
\begin{equation}
\begin{aligned}
    \mathcal P_v(p;\eta_1,\eta_2)
    &=
    e^{-p^2d_{D,12}^2}\mathcal P_v^0(pL_f),
    \\
    d_{D,12}^2
    &\equiv
    d_D^2(\eta_*,\eta_1)+d_D^2(\eta_*,\eta_2).
\end{aligned}
    \label{eq:app_diss_velocity_spectrum}
\end{equation}
At equal times, this reduces to
$\mathcal P_v(p;\eta,\eta)=
e^{-2p^2d_D^2(\eta_*,\eta)}\mathcal P_v^0(pL_f)$, consistent with the
squared amplitude damping in Eq.~\eqref{eq:app_damped_velocity_mode}.

Combining the damped velocity spectrum with the TT-projected stress, we obtain
the physical dissipative UETC
\begin{equation}
\begin{aligned}
    \Pi_{\rm diss}(q,\eta_1,\eta_2)
    &=
    \frac{a_s^8L_f^3}{a^4(\eta_1)a^4(\eta_2)}
    \left[\bar w_{f,s}\bar U_f^2\right]^2
    \\
    &\quad\times
    \tilde\Pi_{\rm diss}(Q,q\eta_1,q\eta_2).
\end{aligned}
    \label{eq:app_dimensionless_diss_uetc}
\end{equation}
The dimensionless momentum convolution is
\begin{equation}
\begin{aligned}
    \tilde\Pi_{\rm diss}(Q,q\eta_1,q\eta_2)
    &=
    \frac{\pi}{2\bar U_f^4}
    \int d^3\tilde p\,
    \frac{(1-\mu^2)^2}{\tilde p\,\tilde{\bar p}^{5}}
    \mathcal P_v^0(\tilde p)
    \mathcal P_v^0(\tilde{\bar p})
    \\
    &\quad\times
    \cos[c_s p(\eta_1-\eta_2)]
    \cos[c_s\bar p(\eta_1-\eta_2)]
    \\
    &\quad\times
    \exp\left[-(\tilde p^2+\tilde{\bar p}^{\,2})
    \frac{d_{D,12}^2}{L_f^2}\right],
\end{aligned}
    \label{eq:app_full_diss_uetc}
\end{equation}
where the dimensionless momenta and angular variable are defined above. The
two internal velocity spectra produce the two terms in the damping exponent.
Accordingly, setting $d_{D,12}^2=0$ recovers the nondissipative kernel
$\tilde\Pi_0$ in Eq.~\eqref{eq:app_sound_shell_uetc}.

To separate the central-time and momentum dependence, we first apply the
frozen-damping approximation
$D_a(\eta)\simeq D_{a,*}\equiv D_a(\eta_*)$. Equations~\eqref{eq:app_dD_definition}
and \eqref{eq:app_yplus_yminus_eta} then give
\begin{equation}
\begin{aligned}
    d_{D,12}^2
    &=
    D_{a,*}(\eta_1+\eta_2-2\eta_*)
    \\
    &=
    2D_{a,*}(\eta_+-\eta_*)
    \\
    &=
    \frac{2D_{a,*}}{a_sH_s}(y_+-y_{+,*}),
\end{aligned}
    \label{eq:app_dD_yplus}
\end{equation}
where
\begin{equation}
    y_{+,*}
    \equiv
    \frac{2}{1+3w}+a_sH_s(\eta_*-\eta_s).
    \label{eq:app_yplus_star}
\end{equation}
Defining
\begin{equation}
    \hat\delta_D
    \equiv
    \frac{2D_{a,*}}{a_sH_sL_f^2},
    \label{eq:app_hat_delta}
\end{equation}
the damping factor in Eq.~\eqref{eq:app_full_diss_uetc} becomes
\begin{equation}
    \exp\left[-(\tilde p^2+\tilde{\bar p}^{\,2})
    \hat\delta_D(y_+-y_{+,*})\right].
    \label{eq:app_frozen_damping_exponent}
\end{equation}
Unlike the nondissipative UETC, this damping factor couples the central time
$y_+$ to the internal momentum combination
$\tilde p^2+\tilde{\bar p}^{\,2}$. It therefore cannot be taken outside the
momentum integral, and the previous factorization no longer applies directly.

We recover the separation with the characteristic-momentum approximation,
which evaluates the momentum factor in the damping exponent at the
characteristic momenta of the nondissipative source integrand,
\begin{equation}
    \tilde p^2+\tilde{\bar p}^{\,2}
    \longrightarrow
    S_*(Q),
    \label{eq:app_characteristic_replacement}
\end{equation}
while leaving the remaining nondissipative UETC unchanged. The dissipative
kernel then becomes
\begin{equation}
    \tilde\Pi_{\rm diss}(Q,q\eta_1,q\eta_2)
    \simeq
    e^{-\delta_D(q)(y_+-y_{+,*})}
    \tilde\Pi_0(Q,\tilde q\,y_-),
    \label{eq:app_piUETCindelta}
\end{equation}
where
\begin{equation}
    \delta_D(q)
    \equiv
    \hat\delta_D S_*(Q),
    \qquad
    S_*(Q)=\tilde p_*^2(Q)+\tilde{\bar p}_*^2(Q).
    \label{eq:app_deltaD}
\end{equation}
This replacement restores the same separation of time scales as in the
nondissipative case, giving
\begin{equation}
    \mathcal P_{\rm GW,diss}(y,Q)
    =
    \tilde P_{\rm GW}(Q)
    \Upsilon_{\rm diss}(y,q;w)y^{3w-1},
    \label{eq:app_diss_factorized}
\end{equation}
with
\begin{equation}
    \Upsilon_{\rm diss}(y,q;w)
    =
    c_w^{\gamma_w}
    \int_{c_w}^{c_wy^{\alpha_w}}
    dy_+
    y_+^{-\gamma_w}
    e^{-\delta_D(q)(y_+-y_{+,*})}.
    \label{eq:app_upsilon_diss_integral}
\end{equation}
Thus damping modifies only the central-time growth factor, while the
source-profile dependence is retained in $\tilde P_{\rm GW}(Q)$.

The remaining central-time integral can be evaluated in terms of the upper
incomplete gamma function. We write
\begin{equation}
    e^{-\delta_D(q)(y_+-y_{+,*})}
    =
    e^{\delta_D(q)y_{+,*}}e^{-\delta_D(q)y_+},
\end{equation}
and introduce
\begin{equation}
    s_w\equiv1-\gamma_w=\frac{3w-3}{1+3w},
    \qquad
    x_q\equiv c_w\delta_D(q).
    \label{eq:app_sw_xq}
\end{equation}
Applying the identity
\begin{equation}
    \int x^{\nu-1}e^{-\lambda x}dx
    =
    -\lambda^{-\nu}\Gamma(\nu,\lambda x),
    \label{eq:app_gamma_identity}
\end{equation}
gives
\begin{equation}
    \Upsilon_{\rm diss}(y,q;w)
    =
    \mathcal A_w(q)
    e^{\delta_D(q)y_{+,*}}
    \left[
    \Gamma(s_w,x_q)
    -
    \Gamma(s_w,x_qy^{\alpha_w})
    \right],
    \label{eq:app_upsilon_closed}
\end{equation}
where
\begin{equation}
    \mathcal A_w(q)
    \equiv
    c_w^{\gamma_w}\delta_D(q)^{-s_w}.
    \label{eq:app_Aw}
\end{equation}
This is the closed form used in Eq.~\eqref{eq:Upsilondissipation}  of the main text. Since
$\eta_*=\eta_s$, one has $y_{+,*}=c_w$. We therefore introduce
\begin{equation}
    t\equiv y_+-c_w,
    \qquad
    \Delta_y\equiv c_w\left(y^{\alpha_w}-1\right),
    \label{eq:app_delta_y}
\end{equation}
which gives the equivalent representation
\begin{equation}
    \Upsilon_{\rm diss}(y,q;w)
    =
    \int_0^{\Delta_y}dt\,
    \left(1+\frac{t}{c_w}\right)^{-\gamma_w}
    e^{-\delta_D(q)t}.
    \label{eq:app_upsilon_shifted}
\end{equation}
This form also makes the limiting behavior clear. Taking the zero-damping
limit in Eq.~\eqref{eq:app_upsilon_shifted} gives
\begin{equation}
    \lim_{\delta_D\to0}
    \Upsilon_{\rm diss}(y,q;w)
    =
    c_w^{\gamma_w}
    \int_{c_w}^{c_wy^{\alpha_w}}
    dy_+\,y_+^{-\gamma_w}
    =
    \Upsilon(y;w).
    \label{eq:app_zero_damping_integral}
\end{equation}
Unlike a term-by-term expansion of the incomplete gamma functions, the
integral form remains regular for negative integer $s_w$, including $s_w=-1$
in radiation domination.

Within the characteristic-momentum approximation, the same source spectrum
$\tilde P_{\rm GW}(Q)$ appears in the dissipative and nondissipative spectra
and therefore cancels in their ratio:
\begin{equation}
    \mathcal R_D(y,q;w)
    \equiv
    \frac{\mathcal P_{\rm GW,diss}(y,Q)}
         {\mathcal P_{\rm GW}^{0}(y,Q)}
    =
    \frac{\Upsilon_{\rm diss}(y,q;w)}{\Upsilon(y;w)}.
    \label{eq:app_ratio_definition}
\end{equation}
Equation~\eqref{eq:app_zero_damping_integral} immediately gives
$\mathcal R_D\to1$ as $\delta_D\to0$.

We next consider the opposite limit. In the strong-damping regime, defined by
$c_w\delta_D(q)\gg1$ and $\delta_D(q)\Delta_y\gg1$, the exponential in
Eq.~\eqref{eq:app_upsilon_shifted} confines the integral to $t\simeq0$.
Expanding the remaining factor about the lower endpoint gives
\begin{equation}
    \left(1+\frac{t}{c_w}\right)^{-\gamma_w}
    =
    1-\frac{\gamma_w}{c_w}t
    +\frac{\gamma_w(\gamma_w+1)}{2c_w^2}t^2
    +O(t^3).
    \label{eq:app_endpoint_expansion}
\end{equation}
The upper limit may then be extended to infinity with only an exponentially
small error. Integrating term by term yields
\begin{equation}
\begin{aligned}
    \Upsilon_{\rm diss}(y,q;w)
    =
    &\frac{1}{\delta_D(q)}
    -\frac{\gamma_w}{c_w\delta_D^2(q)}
    +\frac{\gamma_w(\gamma_w+1)}{c_w^2\delta_D^3(q)}
    +O\!\left(\delta_D^{-4}(q)\right)
    \\
    &\quad+O\!\left(\frac{e^{-\delta_D(q)\Delta_y}}
    {\delta_D(q)}\right).
\end{aligned}
    \label{eq:app_upsilon_strong_expansion}
\end{equation}
Using $\gamma_w/c_w=2$, this expansion becomes
\begin{equation}
    \Upsilon_{\rm diss}(y,q;w)
    =
    \frac{1}{\delta_D(q)}
    \left[
    1-\frac{2}{\delta_D(q)}
    +\frac{5+3w}{\delta_D^2(q)}
    +O\!\left(\delta_D^{-3}(q)\right)
    \right].
    \label{eq:app_upsilon_strong_compact}
\end{equation}
For $w<1$, the nondissipative growth factor approaches the constant
\begin{equation}
    \Upsilon_\infty(w)
    =
    \frac{c_w}{\gamma_w-1}
    =
    \frac{2}{3(1-w)}.
    \label{eq:app_upsilon_nondiss_late}
\end{equation}
Substituting Eqs.~\eqref{eq:app_upsilon_strong_compact} and
\eqref{eq:app_upsilon_nondiss_late} into
Eq.~\eqref{eq:app_ratio_definition} gives
\begin{equation}
    \mathcal R_D^\infty(q;w)
    =
    \frac{3(1-w)}{2\delta_D(q)}
    \left[
    1-\frac{2}{\delta_D(q)}
    +O\!\left(\delta_D^{-2}(q)\right)
    \right],
    \qquad w<1.
    \label{eq:app_ratio_strong}
\end{equation}
The leading term is the strong-damping result quoted in the main text. For
$w\geq1$, however, the nondissipative growth factor does not approach a
constant, and the ratio retains an explicit time dependence. At $w=1$,
$\Upsilon(y;1)=\ln y$, so
\begin{equation}
    \mathcal R_D(y,q;1)
    \simeq
    \frac{1}{\delta_D(q)\ln y}
    \left[
    1-\frac{2}{\delta_D(q)}
    +O\!\left(\delta_D^{-2}(q)\right)
    \right].
    \label{eq:app_ratio_kination}
\end{equation}
For $w>1$, $\Upsilon(y;w)\simeq
2y^{3(w-1)/2}/[3(w-1)]$, while the dissipative factor remains finite. It
follows that
\begin{equation}
    \mathcal R_D(y,q;w)
    \simeq
    \frac{3(w-1)}{2\delta_D(q)}
    y^{-3(w-1)/2},
    \qquad w>1.
    \label{eq:app_ratio_stiff}
\end{equation}
The omitted relative corrections are $O(\delta_D^{-1})$ and
$O(y^{-3(w-1)/2})$.

\section{Numerical validation of the characteristic-momentum approximation}
\label{appendix:characteristicmomentum}

The compact form of $\Upsilon_{\rm diss}$ derived in the preceding section
rests on replacing the pair-dependent momentum factor $S$ with $S_*(Q)$. We now
quantify the accuracy of this closure directly at the level of the UETC,
before performing the GW time integrals. The comparison proceeds in two
stages: the equal-time test isolates the dissipative momentum reweighting,
whereas the unequal-time test restores the full acoustic phase.

\textit{Equal-time test}.---At fixed $Q$ and $y_+$, we define
\begin{equation}
    S(\tilde{\mathbf p},\mathbf Q)
    \equiv
    \tilde p^2+\tilde{\bar p}^{\,2},
    \qquad
    \lambda\equiv\hat\delta_D(y_+-y_{+,*}),
    \qquad
    \tilde{\bar p}=|\tilde{\mathbf p}-\mathbf Q|.
    \label{eq:app_validation_S_lambda}
\end{equation}
Apart from the common normalization in
Eq.~\eqref{eq:app_sound_shell_uetc}, the nondissipative equal-time momentum
weight is
\begin{equation}
    \mathcal K_Q(\tilde{\mathbf p})
    \equiv
    \frac{(1-\mu^2)^2}
         {\tilde p\,\tilde{\bar p}^{\,5}}
    \mathcal P_v^0(\tilde p)
    \mathcal P_v^0(\tilde{\bar p}),
    \qquad
    \mu=\hat{\mathbf q}\cdot\hat{\mathbf p}.
    \label{eq:app_validation_kernel}
\end{equation}
Because $\mathcal K_Q$ is nonnegative, the equal-time comparison isolates the
dissipative reweighting without acoustic phase cancellations.

For reproducibility, we determine the characteristic momenta from the
integration density in $(\tilde p,\tilde{\bar p})$. Using
$d^3\tilde p=2\pi\tilde p^2d\tilde p\,d\mu$ and
$d\mu=-\tilde{\bar p}\,d\tilde{\bar p}/(\tilde pQ)$ gives
\begin{equation}
    \mathscr K_Q(\tilde p,\tilde{\bar p})
    =
    \frac{2\pi}{Q}
    \frac{(1-\mu^2)^2}{\tilde{\bar p}^{\,4}}
    \mathcal P_v^0(\tilde p)
    \mathcal P_v^0(\tilde{\bar p}),
    \label{eq:app_peak_density}
\end{equation}
on the triangle domain
$|\tilde p-\tilde{\bar p}|\leq Q\leq\tilde p+\tilde{\bar p}$, where
$\mu=(\tilde p^2+Q^2-\tilde{\bar p}^{\,2})/(2\tilde pQ)$. We identify the
maximum $(\tilde p_*,\tilde{\bar p}_*)$ of $\mathscr K_Q$ and define the
corresponding characteristic damping scale as
\begin{equation}
    S_*(Q)=\tilde p_*^2(Q)+\tilde{\bar p}_*^2(Q).
    \label{eq:app_validation_Sstar}
\end{equation}
The full convolution and its characteristic-momentum approximation are then
\begin{align}
    \Pi_{\rm diss}^{\rm full}(Q,\lambda)
    &=
    \int d^3\tilde p\,
    \mathcal K_Q e^{-\lambda S},
    \\
    \Pi_{\rm diss}^{\rm cm}(Q,\lambda)
    &=
    e^{-\lambda S_*(Q)}
    \int d^3\tilde p\,\mathcal K_Q.
    \label{eq:app_cm_equal_time}
\end{align}
We use
\begin{equation}
    \Pi_0(Q)
    \equiv
    \Pi_{\rm diss}^{\rm full}(Q,0)
    =
    \int d^3\tilde p\,\mathcal K_Q
    \label{eq:app_validation_Pi0}
\end{equation}
to denote the nondissipative equal-time convolution with the same omitted
normalization. Throughout the scan, $S_*(Q)$ is fixed by the nondissipative
equal-time density and is not readjusted as $\lambda$ varies. The comparison
therefore probes only the error incurred by evaluating the
momentum-dependent damping exponent at this fixed scale.

\begin{figure}[htb]
    \centering
    \includegraphics[width=0.62\textwidth]{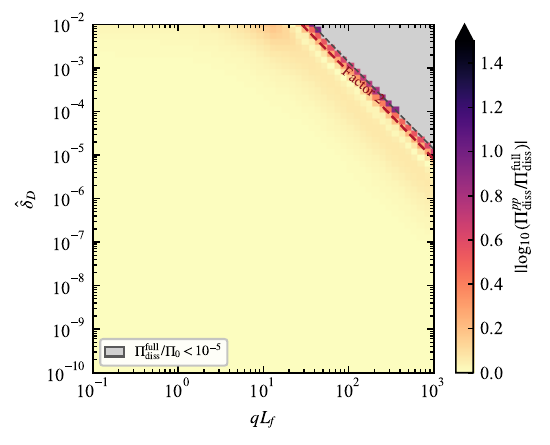}
    \caption{Equal-time validation of the characteristic-momentum
    approximation. The heat map shows
    $|\log_{10}(\Pi_{\rm diss}^{\rm cm}/\Pi_{\rm diss}^{\rm full})|$ in the
    $(Q,\hat\delta_D)$ plane for $y_{+,*}=1$, $y_+=2$, and $y_-=0$. The red
    dashed contour marks a factor-of-two deviation from the full momentum
    convolution. The gray region denotes
    $\Pi_{\rm diss}^{\rm full}/\Pi_0<10^{-5}$, where the dissipative kernel is
    already strongly suppressed relative to its nondissipative value.}
    \label{fig:app_effective_momentum_validation}
\end{figure}

We quantify this error at each point in the $(Q,\hat\delta_D)$ plane by
\begin{equation}
    \Delta_{\rm cm}(Q,\hat\delta_D)
    \equiv
    \left|
    \log_{10}
    \frac{\Pi_{\rm diss}^{\rm cm}}
         {\Pi_{\rm diss}^{\rm full}}
    \right|.
    \label{eq:app_cm_validation_metric}
\end{equation}
Figure~\ref{fig:app_effective_momentum_validation} shows this comparison for
$y_{+,*}=1$ and $y_+=2$. We scan 54 logarithmically spaced values of $Q$ over
$[10^{-1},10^3]$ and 58 values of $\hat\delta_D$ over
$[10^{-10},10^{-2}]$. The velocity spectrum is taken from the sound-shell
model benchmark of Fig.~12 in Ref.~\cite{Guo:2020grp}, with $\alpha=0.0046$
and $v_w=0.92$. The momentum integrals are evaluated with composite
Gauss--Legendre quadrature in $(\log\tilde{\bar p},\tilde p)$. Numerical
stability is checked at representative high-$Q$ points, where extending the
integration domain and increasing the quadrature order changes the relevant
ratios by less than $5\times10^{-4}$.

The approximation tracks the full equal-time convolution across most of the
scanned plane. Deviations larger than one order of magnitude appear only
after the full kernel has fallen below $10^{-5}$ of its nondissipative value.
The factor-of-two contour reaches only slightly beyond this region; even at
its least-suppressed point,
$\Pi_{\rm diss}^{\rm full}/\Pi_0\simeq5.7\times10^{-4}$. The mismatch is thus
concentrated in the strongly damped high-frequency corner.

\textit{Unequal-time test}.---The equal-time test cleanly isolates the
dissipative envelope, but GW production probes the UETC at unequal times. For
$y_-\ne0$, the momentum convolution also contains the acoustic phase
\begin{equation*}
    C(\tilde p,\tilde{\bar p};y_-)
    =
    \cos\!\left[
        \frac{c_s\tilde p}{\beta_cL_f}
        \frac{\beta_c}{a_sH_s}y_-
    \right]
    \cos\!\left[
        \frac{c_s\tilde{\bar p}}{\beta_cL_f}
        \frac{\beta_c}{a_sH_s}y_-
    \right],
    \qquad
    \tilde{\bar p}=|\tilde{\mathbf p}-\mathbf Q|.
\end{equation*}
Either cosine can change sign across the momentum domain, producing partial
cancellations in the convolution. Since dissipation reweights the positive
and negative contributions differently, these cancellations provide a
distinct test of the approximation beyond the equal-time limit.

\begin{samepage}
We therefore repeat the comparison with the sign of $C$ retained in both
calculations. The full convolution and its characteristic-momentum
approximation are
\begin{align*}
    \Pi_{\rm diss}^{\rm full}
    &=\int d^3\tilde p\,
      \mathcal K_Q C\,e^{-\lambda S},
    \\
    \Pi_{\rm diss}^{\rm cm}
    &=e^{-\lambda S_*(Q)}
      \int d^3\tilde p\,\mathcal K_Q C,
    \qquad
    \lambda=\hat\delta_D(y_+-y_{+,*}).
\end{align*}
\end{samepage}
The equal-time scale $S_*(Q)$ is kept fixed at nonzero $y_-$, so any additional
discrepancy directly measures the effect of relative-time phase
cancellations.

To respect the hierarchy $y_+\gg|y_-|$, we take
$y_{+,*}=1$, $y_+=\{2,5,10\}$, and
$|y_-|=\{0,0.002,0.01\}$, for which the smallest nonzero ratio
$y_+/|y_-|$ is 200. We also set $c_s=1/\sqrt{3}$,
$\beta_c/(a_sH_s)=100$, and
$\beta_cL_f=(8\pi)^{1/3}v_w$ with $v_w=0.3$, while keeping the same
dimensionless velocity spectrum as in the equal-time test. Here $v_w=0.3$
enters only the phase conversion and tests a more rapidly varying acoustic
phase. For each pair $(y_+,|y_-|)$, we scan 32 values of $Q$ over
$[10^{-1},10^3]$ and 42 values of $\hat\delta_D$ over
$[10^{-10},10^{-2}]$. Because the ratio becomes ill-conditioned when the
signed full convolution approaches zero, we retain only points satisfying
$|\Pi_{\rm diss}^{\rm full}|/\Pi_{0,\rm abs}\geq10^{-5}$, where
$\Pi_{0,\rm abs}\equiv\int d^3\tilde p\,|\mathcal K_QC|$.

\begin{figure}[b]
    \centering
    \includegraphics[width=0.90\textwidth]{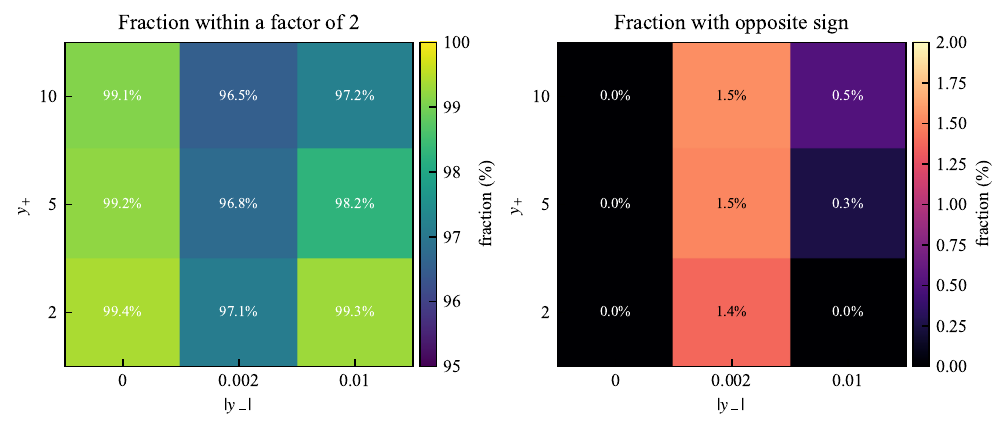}
    \caption{Signed unequal-time validation of the characteristic-momentum
    approximation. Each cell
    summarizes a $32\times42$ scan over $(Q,\hat\delta_D)$ at one fixed pair of
    $y_+$ and $|y_-|$. The signed acoustic phase is retained. Only points with
    $|\Pi_{\rm diss}^{\rm full}|/\Pi_{0,\rm abs}\geq10^{-5}$ are included. The
    left panel gives the percentage with the same sign and a magnitude that
    agrees within a factor of two. The right panel gives the percentage with
    the opposite sign.}
    \label{fig:app_characteristic_momentum_broad_validation}
\end{figure}

Figure~\ref{fig:app_characteristic_momentum_broad_validation} summarizes the nine
slices. Depending on $(y_+,|y_-|)$, $96.5\%$--$99.4\%$ of the retained points
have the same sign and agree in magnitude within a factor of two; the
opposite-sign fraction remains about $1.5\%$ or below. Larger magnitude errors
occur in a small subset, including cancellation-sensitive points. Thus the test
supports the approximation over most of the retained domain, without making
a pointwise claim near zeros of the signed convolution.

Together, the equal- and unequal-time tests quantify the domain in which the
characteristic-momentum closure reproduces the full UETC momentum convolution
over the tested benchmark domain. Its role is limited to obtaining the compact
finite-time growth factor $\Upsilon_{\rm diss}$. The asymptotic high-frequency
scaling does not depend on this closure: in the following section, we restore
the full pair-momentum dependence and derive the strong-damping result
directly.

\clearpage

\section{High-frequency scaling from the full momentum convolution}
\label{appendix:fullq2}

To establish the high-frequency scaling without invoking the
characteristic-momentum approximation, we evaluate
Eq.~\eqref{eq:app_gw_double_time} with the full internal momentum dependence
of the dissipative UETC. The Green-function combination is defined
by~\cite{Xiao:2024rsj}
\begin{equation}
    \frac{\partial G(\tilde y,\tilde y_1)}{\partial\tilde y}
    \frac{\partial G(\tilde y,\tilde y_2)}{\partial\tilde y}
    =
    \frac{\mathcal G(\tilde y,\tilde y_1,\tilde y_2)}
    {\tilde y^{\,3-3w}},
    \label{eq:app_green_combination_definition}
\end{equation}
where $G$ is the dimensionless retarded Green function. For
$\tilde q\gg1$ and $y_+\gg|y_-|$, its large-argument form separates into a
slowly varying envelope and two rapidly oscillating phases. Averaging over
their common phase gives
\begin{equation}
    \overline{
    \cos[q(\eta-\eta_1)]\cos[q(\eta-\eta_2)]}
    =
    \frac12\cos[q(\eta_1-\eta_2)]
    =
    \frac12\cos(\tilde qy_-).
    \label{eq:app_green_phase_average}
\end{equation}
The leading nonoscillatory contribution is therefore
\begin{equation}
    \frac{\mathcal G(\tilde y,\tilde y_1,\tilde y_2)}
    {\tilde q^2
    y_1^{3(1+w)/2}y_2^{3(1+w)/2}}
    \simeq
    \frac12 c_w^{\gamma_w}y_+^{-\gamma_w}
    \cos(\tilde qy_-).
    \label{eq:app_green_relative_phase}
\end{equation}
This result holds for general constant $w$ and reduces to the familiar
radiation-dominated form. Dissipation changes the slowly varying $y_+$
envelope but leaves the relative-time phase $\cos(\tilde qy_-)$ unchanged.

Substituting Eq.~\eqref{eq:app_full_diss_uetc} into
Eq.~\eqref{eq:app_gw_double_time}, and taking $\eta_*=\eta_s$ so that
$y_{+,*}=c_w$, gives
\begin{equation}
\begin{aligned}
    \mathcal P_{\rm GW,diss}^{\rm full}(y,Q)
    \simeq{}&
    \mathcal N(y,Q)
    \int d^3\tilde p\,\mathcal K_Q(\tilde{\mathbf p})
    \\
    &\times\int_{\mathcal D}dy_+dy_-\,
    c_w^{\gamma_w}y_+^{-\gamma_w}
    e^{-\hat\delta_DS(y_+-c_w)}\mathcal K_-(y_-),
\end{aligned}
    \label{eq:app_full_substitution}
\end{equation}
where
\begin{equation}
\begin{gathered}
    \mathcal K_Q(\tilde{\mathbf p})
    =
    \frac{(1-\mu^2)^2}{\tilde p\,\tilde{\bar p}^{5}}
    \mathcal P_v^0(\tilde p)\mathcal P_v^0(\tilde{\bar p}),
    \qquad
    S=\tilde p^2+\tilde{\bar p}^{\,2},
    \\
    \mathcal K_-(y_-)
    =
    \cos(\tilde qy_-)
    \cos\!\left(\frac{c_s\tilde p}{Q}\tilde qy_-\right)
    \cos\!\left(\frac{c_s\tilde{\bar p}}{Q}\tilde qy_-\right).
\end{gathered}
    \label{eq:app_three_cosine_kernel}
\end{equation}
The first cosine is the relative-time phase of the propagating GW, while the
other two are carried by the internal sound modes. Their product determines
whether the source remains phase matched with the tensor mode. The common
normalization is
\begin{equation}
    \mathcal N(y,Q)
    =
    \frac{(16\pi G\bar w_{f,s})^2Q^3}
    {96\pi H^2(y)H_s^2y^4}.
    \label{eq:app_full_common_normalization}
\end{equation}

The scale hierarchy specified below makes the relative-time oscillation scale
much shorter than the evolution scale of the envelope. Boundary effects are
then subleading, and the double-time integral separates. Its central-time
part is
\begin{equation}
\begin{aligned}
    \mathcal I_y(\lambda)
    &\equiv
    c_w^{\gamma_w}
    \int_{c_w}^{c_wy^{\alpha_w}}dy_+\,
    y_+^{-\gamma_w}e^{-\lambda(y_+-c_w)}
    \\
    &=
    \int_0^{\Delta_y}dt\,
    \left(1+\frac{t}{c_w}\right)^{-\gamma_w}e^{-\lambda t},
    \qquad
    \lambda=\hat\delta_DS.
\end{aligned}
    \label{eq:app_full_time_kernel_definition}
\end{equation}

The three phases in $\mathcal K_-$ generate four sum- and
difference-frequency branches. Applying the product-to-sum identity and
$\int_{-\infty}^{\infty}dy\,\cos(\omega y)
=2\pi\delta_{\rm Dirac}(\omega)$, the long-time integral selects their
resonant support. For $Q>0$ and $0<c_s<1$, the difference-frequency branches
are excluded by $|\tilde p-\tilde{\bar p}|\leq Q$, while the opposite
sum-frequency branch has no support. The remaining resonance is
\begin{equation}
\begin{aligned}
    \mathcal J_Q(\tilde p,\tilde{\bar p})
    &\equiv
    \int_{-\infty}^{\infty}dy_-\,\mathcal K_-(y_-)
    \\
    &=
    \frac{\pi Q}{2\tilde q}\,
    \delta_{\rm Dirac}
    \!\left[Q-c_s(\tilde p+\tilde{\bar p})\right].
\end{aligned}
    \label{eq:app_resonance_kernel}
\end{equation}

Defining the resonant weight
\begin{equation}
    \mathcal W_Q(\tilde{\mathbf p})
    \equiv
    \mathcal K_Q(\tilde{\mathbf p})
    \mathcal J_Q(\tilde p,\tilde{\bar p}),
    \label{eq:app_full_momentum_weight}
\end{equation}
we finally obtain
\begin{equation}
    \mathcal P_{\rm GW,diss}^{\rm full}(y,Q)
    \simeq
    \mathcal N(y,Q)
    \int d^3\tilde p\,
    \mathcal W_Q(\tilde{\mathbf p})
    \mathcal I_y\!\left[\hat\delta_D
    S(\tilde{\mathbf p},\mathbf Q)\right].
    \label{eq:app_full_spectrum_time_kernel}
\end{equation}
At zero damping, $\mathcal I_y(0)=\Upsilon(y;w)$, and
Eq.~\eqref{eq:app_full_spectrum_time_kernel} gives
$\mathcal P_{\rm GW}^{0}\simeq\mathcal N\Upsilon
\int d^3\tilde p\,\mathcal W_Q$.
On the support of $\mathcal W_Q$, the resonance condition is
$\tilde p+\tilde{\bar p}=Q/c_s$, and hence
\begin{equation}
    \frac{Q^2}{2c_s^2}
    \leq S(\tilde{\mathbf p},\mathbf Q)
    \leq\frac{Q^2}{c_s^2}.
    \label{eq:app_resonant_S_bounds}
\end{equation}
To extract a uniform strong-damping limit, the relative-time phase must
oscillate many times within even the shortest damping interval, so that the
resonance is resolved before the envelope changes appreciably. At the same
time, even the longest damping interval must remain short compared with both
the background-evolution scale and the available source duration. Since
$\lambda$ varies across the resonance surface, these requirements become
\begin{equation}
\begin{gathered}
    \lambda_{\min}
    =\dfrac{\hat\delta_DQ^2}{2c_s^2},
    \qquad
    \lambda_{\max}
    =\dfrac{\hat\delta_DQ^2}{c_s^2},
    \\
    \tilde q^{-1}
    \ll\lambda_{\max}^{-1}
    \leq\lambda^{-1}
    \leq\lambda_{\min}^{-1}
    \ll\min(c_w,\Delta_y).
\end{gathered}
    \label{eq:app_uniform_strong_conditions}
\end{equation}
The left inequality controls the resonance and its boundary corrections,
whereas the right inequality localizes the central-time integral. The latter
then admits the strong-damping expansion
\begin{equation}
    \mathcal I_y(\lambda)
    =
    \frac{1}{\lambda}
    \left[1+O(\lambda^{-1})+O(e^{-\lambda\Delta_y})\right].
    \label{eq:app_full_time_kernel_asymptotic}
\end{equation}
Because $\mathcal W_Q$ is nonnegative, the bounds in
Eq.~\eqref{eq:app_resonant_S_bounds} carry directly through the momentum
convolution. The source-profile dependence can therefore be collected into
\begin{equation}
    C_{\mathcal W}(Q)
    \equiv
    Q^2
    \frac{
    \displaystyle\int d^3\tilde p\,
    \mathcal W_Q(\tilde{\mathbf p})
    S^{-1}(\tilde{\mathbf p},\mathbf Q)
    }{
    \displaystyle\int d^3\tilde p\,
    \mathcal W_Q(\tilde{\mathbf p})
    },
    \qquad
    c_s^2\leq C_{\mathcal W}(Q)\leq2c_s^2.
    \label{eq:app_kernel_coefficient}
\end{equation}
Combining this result with Eq.~\eqref{eq:app_full_time_kernel_asymptotic}
gives
\begin{equation}
\begin{aligned}
    \mathcal R_{D,\rm full}(y,Q;w)
    &\equiv
    \frac{\mathcal P_{\rm GW,diss}^{\rm full}(y,Q)}
         {\mathcal P_{\rm GW}^{0}(y,Q)}
    \\
    \simeq
    &\frac{C_{\mathcal W}(Q)}
    {\hat\delta_D\Upsilon(y;w)Q^2}
    \left[
    1+O\!\left((\hat\delta_DQ^2)^{-1}\right)
    +O\!\left(e^{-\hat\delta_DQ^2\Delta_y/(2c_s^2)}\right)
    +O\!\left(\frac{\lambda_{\max}}{\tilde q}\right)
    \right]
    .
\end{aligned}
    \label{eq:app_full_qminus2}
\end{equation}
The source profile thus changes only the bounded coefficient
$C_{\mathcal W}$, while the additional $Q^{-2}$ suppression follows from the
full momentum convolution. When
$\Upsilon(y;w)\to2/[3(1-w)]$ for $w<1$, Eq.~\eqref{eq:app_full_qminus2}
reduces to
\begin{equation}
    \mathcal R_{D,\rm full}^{\rm late}(Q;w)
    \simeq
    \frac{3(1-w)C_{\mathcal W}(Q)}
         {2\hat\delta_D Q^2},
    \qquad w<1.
    \label{eq:app_full_qminus2_late}
\end{equation}
The characteristic-momentum approximation corresponds to replacing
$C_{\mathcal W}(Q)$ by $Q^2/S_*(Q)$: it estimates the amplitude but leaves the
$Q^{-2}=(qL_f)^{-2}$ scaling unchanged.

